# Features of clustering in a supersonic argon jet when using a conical nozzle.

**Yu. S. Doronin, A. A. Tkachenko, V. L. Vakula, G. V. Kamarchuk**

*B. Verkin Institute for Low Temperature Physics and Engineering of NAS of Ukraine,
47 Nauky Ave., Kharkiv, Ukraine
E-mail: doronin@ilt.kharkov.ua*

The article presents an original spectroscopic method for determining the initial stage of the clustering process in supersonic jets. The technique has been tested on a supersonic argon jet excited by electrons at a distance of 30 millimetres from the nozzle outlet, where the clustering process is practically complete, and the influence of secondary processes on the intensity of the observed emissions is significantly suppressed. It is proposed to use the continuum emitted by neutral excited clusters with a wavelength of $\lambda=127$ nm as an indicator of cluster formation in a supersonic argon jet under various flow conditions. Analysis of the temperature dependence of the intensity of this continuum recorded at multiple argon pressures at the nozzle inlet allowed us to establish that the parameters of supersonic jet flow corresponding to the onset of crystallisation are related by the empirical expression $P_0 T_0^{-2.7} =$ const. The calculated value of the constant for argon in the studied range of pressures and temperatures was 0.011.

## 1. Introduction

Supersonic jets of inert and molecular gases in atomic and cluster flow modes are widely used in a range of scientific and technological applications. The use of atomic supersonic jets as targets for high-energy electron scattering, precision spectroscopy, or in laser-plasma experiments allows the generation of plasma to be spatially separated from interaction with the substrate, which provides precise control in material processing and the study of complex plasma physics, including shock waves and jet-environment interactions. At the same time, it is essential to maintain the atomic composition of the supersonic jet, since clustering leads to the formation of undesirable atomic complexes and significant changes in its physical parameters [1-6].

Recent studies of the interaction of intense laser radiation with clustered supersonic jets have demonstrated higher efficiency in forming a controlled hot nanoplasma during laser irradiation of clusters with small volume and density, close to those of a solid body [7,8]. Due to the efficient absorption of laser energy within the finite size of the cluster, intense X-ray radiation, a large number of energetic electrons, and highly charged ions are generated, followed by a Coulomb explosion or hydrodynamic expansion of clusters, which is of great importance for high-energy physics [9,10].

Currently, cluster and ion-cluster beams are widely used for the synthesis of thin, atomically smooth films, the doping of ultrathin layers, and the surface treatment of various materials. The high transition energy density generated by individual cluster impacts can also significantly increase the rate of surface chemical reactions, even at low temperatures. Several published reviews are devoted to the development of cluster technologies and the results they achieve at the macroscopic and nanoscale [11-13].

For a long time, cluster beams have been the subject of numerous fundamental studies, since the properties of clusters differ from those of the atoms or molecules that comprise them, as well as from those of a continuous medium [14-16]. Currently, new methods for studying clusters are being proposed and used; for example, the size and density of clusters in a jet can be determined using Rayleigh light scattering [17, 18]. Despite this, there is still no adequate analytical description of clustering in supersonic jets. This is because the condensation process in supersonic jets involves rapid non-equilibrium thermodynamics, complex gas phase kinetics (cluster nucleation, growth, and fragmentation), and is sensitive to the parameters of the nozzles used and the initial conditions of the flow, which significantly complicates modelling, especially when working with nanoclusters, where classical condensation theory does not work. Experimental diagnostics do not always provide reliable results. Therefore, in many works, authors continue to use the well-known semi-empirical Hagena's parameter [19-20] to describe the clustering process, leading to significant discrepancies in predicted cluster sizes. To fully utilise the potential of supersonic jets, it is necessary to know several key parameters. These include: the nature of the jet flow and the density or concentration of atoms, the pressure and temperature of the gas at the nozzle inlet at which the clustering process begins in the jet, the ratio of monomers and clusters in the jet at different stages of clustering, the mass fraction of condensate, and the average cluster size. Another problem is the reproducibility of results, which depends heavily on the gas-jet parameters, i.e., not only on the atomic density but also on the cluster size and density.

In experiment [22], the authors used a spectroscopic method based on absolute measurements of the VUV radiation flux generated by the excitation of a supersonic argon jet with an electron beam of 1 keV energy to determine the density of the non-condensed atomic component, the condensate fraction, and the cluster concentration in a supersonic argon jet at given flow parameters.

This experiment demonstrates how this method can be used to determine the parameters of a supersonic jet at which the clustering process begins.

**2. Experiment**

The research was conducted on a spectroscopic stand described in detail in [21]. The main components of the stand are: a cluster beam generator, an electron gun, and a radiation detection system. A conical supersonic nozzle with a critical cross-section diameter of 0.34 mm, a diffuser length of 11 mm, a cone angle of $8.6^0$, and a ratio of the outlet cross-section area to the critical cross-section area of 36.7 was used to form the cluster beam. At a distance of 30 mm from the nozzle outlet, the supersonic argon jet was excited by an electron beam with an electron energy of 1 keV and an electron beam current of 20 mA. The average size of clusters at the point of excitation of the jet by electrons was varied using the pressure $P_0$ and temperature $T_0$ of the gas at the nozzle inlet. In the atomic mode of a supersonic argon jet, VUV radiation is mainly generated by resonant transitions between the ground and excited states of atoms and ions. In the cluster mode, as the number and size of clusters in the jet increase, the total VUV radiation flux is determined by the continuum emitted by neutral and charged excimer complexes formed in the clusters.

Figure 1 shows a typical spectrum of radiation from a supersonic argon jet in the wavelength range 90-150 nm at gas pressure at the nozzle inlet $P_0 = 0.1$ MPa and temperature $T_0 = 180$ K. The spectrum contains resonance lines emitted by argon ions Ar II ($\lambda = 92.0$ nm, $\lambda = 93.2$ nm) and atoms Ar I ($\lambda = 104.8$ nm, $\lambda = 106.7$ nm), as well as a cluster continuum with a maximum $\lambda = 109$ nm (W), emitted by excimers $(Ar_2)^*$ from partially vibrationally relaxed states, and cluster continua emitted by neutral $(Ar_2)^*$ at $\lambda = 127$ nm and charged excimer complexes $(Ar_4^+)^*$ at $\lambda = 137$ nm in vibrationally relaxed states [21,22]. At pressure = 0.1 MPa and argon temperature at the nozzle inlet $T_0 > 400$ K, there are no cluster emissions in the spectrum, and only Ar II and Ar I lines are registered, which indicates the atomic composition of the supersonic argon jet. Under these conditions, the density of argon atoms in the studied region of the jet is $n = 5.1 \cdot 10^{15}$ cm$^{-3}$ [22].

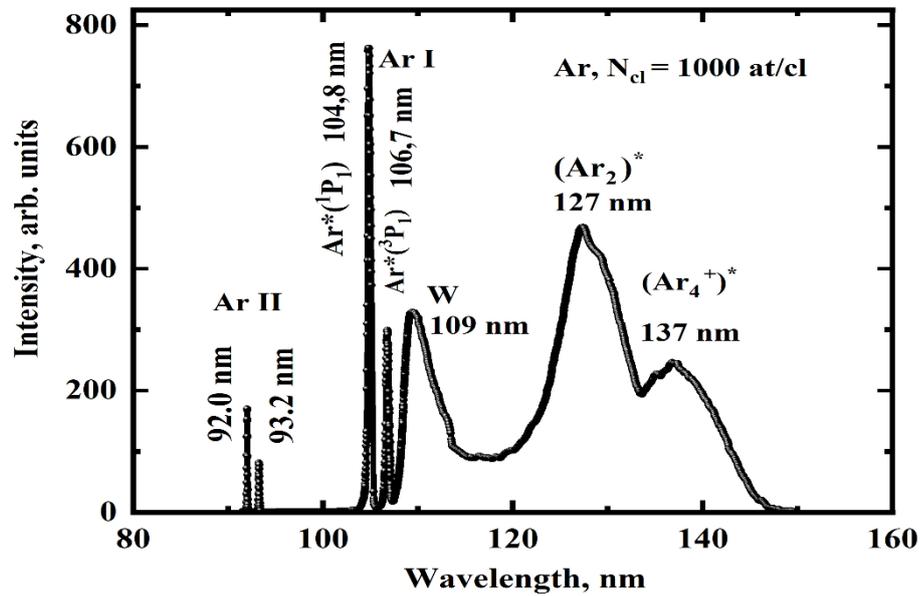

Fig. 1. The spectrum of radiation from a supersonic argon jet, an electron energy of 1 keV, an electron beam current of 20 mA, and a gas pressure and temperature at the nozzle inlet of $P_0 = 0.1$ MPa and $T_0 = 180$ K.

### 3. Results and discussion

At a distance of 30 mm from the nozzle outlet section, with an atom concentration in the jet of approximately $10^{15}$ cm$^{-3}$, the formation of excited argon excimer molecules via triple collisions, characteristic of gaseous argon, is significantly suppressed. Consequently, the appearance of the λ=127 nm continuum in the spectrum is clearly associated with the formation of excimer-type emitting centres ($Ar_2^*$) in neutral argon clusters [21,22].

Fig. 2 shows the temperature dependence of the intensity of the continuum $\lambda = 127$ nm emitted by molecular-type centres ($Ar_2^*$) formed in clusters at several values of $P_0$.

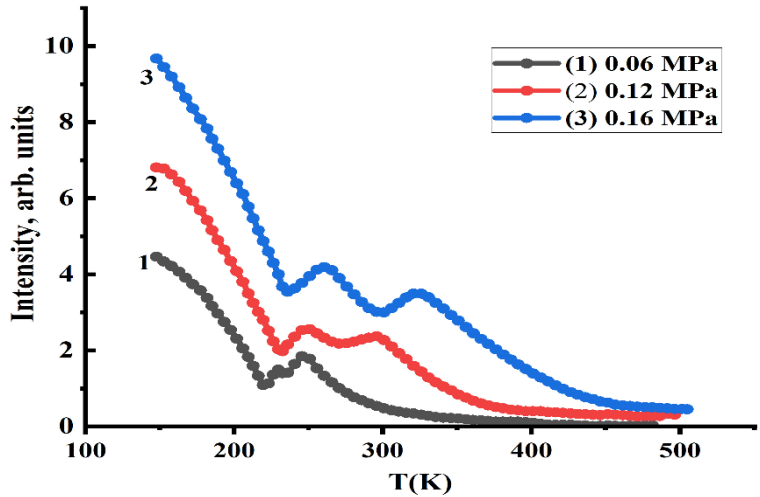

Fig. 2 The temperature dependencies of the intensity of the continuum $\lambda=127$ nm in relative units, at different pressures $P_0$ of the gas at the nozzle inlet: 1-0.06 MPa; 2-0.12 MPa; 3-0.16 MPa.

For the selected pressure values, the appearance and further increase in continuum intensity are observed at different temperatures.

Previously [22], using the STATGRAPHICS plus software package, we found a strong inverse correlation ($r \approx -1$) between the temperature dependencies of the ion line $\lambda=92.0$ nm and the cluster continuum $\lambda=127$ nm at $P_0 = 0.1$ MPa, i.e., the onset of the clustering process correlates with a decrease in the intensity of the Ar II line ($\lambda=92.0$ nm) (see Figs. 3, 4 [22]). Therefore, at the same pressures as for $\lambda=127$ nm, temperature dependencies were recorded for the $\lambda=92.0$ line (Fig. 3).

Analysis of the obtained temperature dependencies (Fig. 2,3) showed that the process of cluster formation in a supersonic argon jet begins at the following values of $P_0$ at the nozzle inlet: $P_0 = 0.06$ MPa – 320 K; $P_0 = 0.14$ MPa – 400 K; $P_0 = 0.16$ MPa – 470 K. The start of the clustering process is described by the empirical expression $P_0 \text{(Pa)} T_0 \text{(K)}^{-2.7} \approx 0.011$. The dependencies of the intensity of the continuum $\lambda=127$ and the line $\lambda=92.0$ measured at $P_0 = 0.1$ MPa in [22] are described by this expression.

The proposed method provides registration of temperature dependencies for lines and continua with an accuracy of approximately 10%.

To describe the process of cluster formation in supersonic jets, O. Hagena [19, 20]. proposed using an empirical condensation scaling parameter based on the principle of 'corresponding jets'. In general, a value of $\Gamma^* > 200$ corresponds to the initial stage of condensation in a supersonic jet. For the nozzle used in this work, the following calculated temperatures for the onset of clustering ($\Gamma^* > 200$) were obtained according to [19, 20]: $P_0$ = 0.06 MPa – 470K; $P_0$ = 0.1 MPa - 780K; $P_0$ = 0.14 MPa - 1000K; $P_0$ = 0.16 MPa – 1250 K. It should be emphasised that the calculated values of $T_0$ at all given $P_0$ significantly exceed the corresponding values of $T_0$ determined experimentally in this work.

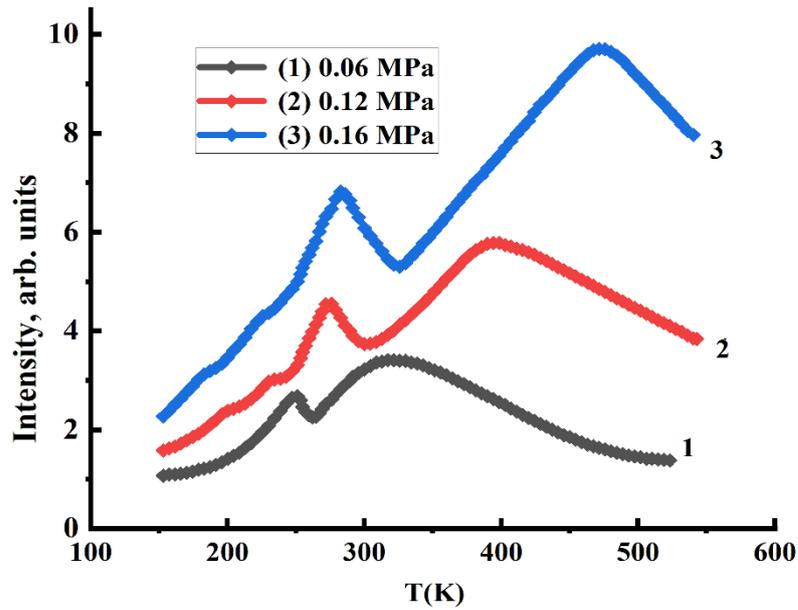

Fig. 3 The temperature dependencies of the intensity of the λ=92 nm line in relative units at different pressures $P_0$ of the gas at the nozzle inlet: 1- 0.06 MPa; 2 - 0.12 MPa; 3 - 0.16 MPa.

The results obtained show that the clustering process in a conical supersonic nozzle begins at significantly lower temperatures than those calculated using the condensation scaling parameter, $\Gamma^*$, introduced by O. Hagena and co-authors. The exponent in the expression $P_0$ (Pa)$T_0$ (K)$^{-2.7} \approx 0.011$, equal to (-2.7), indicates that cluster formation in a supersonic conical nozzle occurs much faster than in sonic nozzles. In addition, the conical nozzle provides higher gas density and lower temperature in the supersonic expansion region, creating ideal conditions for cluster formation (high supersaturation, high collision frequency).

## 4. Conclusions

The paper discusses a spectroscopic method for determining the onset of clustering in a supersonic argon jet formed by a conical nozzle. Analysis of the obtained temperature dependencies of the intensity $I(T_0)$ of the argon cluster continuum with a maximum $\lambda$ = 127 nm and atomic ions line $\lambda$ = 92.0 allowed us to determine the threshold temperature for the appearance of the $\lambda$ = 127 nm continuum in the spectrum at different values of $P_0$, which indicates the beginning of the clustering process in the argon jet. In this case, the pressure $P_0$ and temperature $T_0$ of argon at the nozzle inlet are related by the empirical dependence $P_0$ (Pa) $T_0$ (K) $^{-2.7} \approx 0.011$. Hagena's scaling parameter yields significantly higher $T_0$ values, enabling quick estimates of the possible range of average cluster sizes in a supersonic jet at given flow parameters. For the correct interpretation of the results of experimental and applied work using cluster beams, new methods are needed that allow the size and concentration of clusters at the point of application to be obtained with greater accuracy. The spectroscopic method tested in this work can be applied to supersonic jets of other rare gases and used to correct the results obtained using Hagena's scaling parameter.